\documentstyle[floats,aps,prb,epsfig,eqsecnum,amsmath,amssymb]{revtex}
\newif\ifbig

\bigtrue
\bigfalse

\ifbig
        
\fi

\newcommand{\Name}[1]
{
#1,
}

\renewcommand{\and}
{
and
}

\newcommand{\REVIEW}[4]
{
#1 {\bf #2}, #4 (#3).
}

\begin{document}
\draft
\date{\today}

\ifbig
{}
\else
\twocolumn[\hsize\textwidth\columnwidth\hsize\csname@twocolumnfalse%
\endcsname
\fi
\title{Collective Dynamics of One-Dimensional  Charge Density Waves}
\author{Andreas Glatz${}^1$, Sanjay Kumar${}^2$ and Mai Suan Li${}^{1,3}$}

\address{${}^1$Institut f\"ur Theoretische Physik, Universit\"at zu K\"oln, Z\"ulpicher Str. 77, D-50937 K\"oln, Germany\\ 
        ${}^2$Department of Physics, Banaras Hindu University, Varanasi 221005, India\\
        ${}^3$Institute of Physics, Al. Lotnikow 32/46, 02-668 Warsaw, Poland}

\maketitle
\begin{abstract}
The effect of disorder on the static and dynamic behaviour of one-dimensional
charge density waves at low temperatures is studied by analytical and
numerical approaches. In the low temperature region
the spatial behaviour of the phase-phase correlation function is dominated
by disorder but the roughness exponent remains the same as in
the pure case.
Contrary to high dimensional systems the dependence of
the creep velocity on the electric field is described by an analytic
function.
\end{abstract}

\vspace{1cm}


\ifbig
{}
\else
]
\fi


\section{Introduction}\label{sec:intro}

The collective dynamics of condensed modulated structures like
charge (or spin) density waves  \cite{Gruener88,Gruener94,Brazovski99},
flux line systems \cite{Blatter,NatSchei} and Wigner crystals
\cite{Brazovski99} in random environments has been the subject of detailed
investigation for more than 20 years. In systems with dimension $d>2$
the collective creep of these structures is determined by
a zero temperature disorder fixed point resulting in a non-analytic
current-voltage relation with zero linear resistivity \cite{Natter90}.
In $d=2$ dimensions this fixed point is extended to a fixed line which
terminates at the glass transition temperature $T_g$ \cite{CardyOst}.
The high temperature phase $T>T_g$ is characterized by a
fixed point of zero disorder and a power law decay of the density correlation function.
The dynamic behaviour is Ohmic. In the low temperature phase $T<T_g$,
correlations decay slightly faster than a power law and the
linear resistivity still vanishes (for a recent review, see [\onlinecite{NatSchei}]).

It should be noted that quasi-one-dimensional systems
such as $TaS_3$ \cite{Zaitsev93,Zaitsev94} and whiskers \cite{Brasov2000} have been a subject
of intensive experimental research. From the theoretical point of view these systems 
are of special interest because the situation in less than two dimensions ( $d<2$) 
is entirely different compared to higher dimensions. As follows
already from a dimensional continuation of the Cardy-Ostlund flow
equations \cite{CardyOst} to dimensions $d<2$, the glass 
temperature $T<T_g$ is shifted to
$T=0$. Nevertheless there remains a residual trace of the disorder
which is reflected in the low temperature behaviour of spatial
correlations and the dynamics. To be specific we will denote the
temperature where the influence of disorder gradually sets in by $T^{\ast}$.

The pair correlation function is defined as follows
\begin{equation}
  C(x) \; = \; \big<\overline{\big(\varphi(x)-\varphi(0)\big)^2}\big> \; ,
\label{eq.00}
\end{equation}
where $\varphi(x)$ is the phase of the charge density wave (CDW),
the overbar and the bracket denote the disorder and the thermal
averaging,
respectively. For the one-dimensional system one can show that the roughness
exponent is equal to $\frac{1}{2}$, i.e 
\begin{equation}
  C(x) \; = \; A |x| \; .
\label{eq.000}
\end{equation}
At high temperatures the coefficient $A$ is defined by thermal fluctuations
(disorder is irrelevant) and $A \sim T$.
 At strictly zero temperature $A$ 
was calculated (ignoring effects from
plasticity) previously both by Feigel'man \cite{Feigel80} and
Villain and Fernandez \cite{ViFer84}. 
Their results at $T=0$ predict
essentially the same decay of the correlation function as at high
temperatures but with $T$ replaced by $T^*$. 
An expression for $A$ in the crossover region from the high-$T$ 
to the zero-temperature regime remains unknown.

Therefore it is one of the aims of the present paper to study the influence of disorder
in the - so-far not considered - low-temperature region $0<T \lesssim T^{\ast}$ 
on the static behaviour of the CDW's.
Using the lowest order of perturbation theory we have shown that
at low temperatures the 
coefficient $A$ is a linear function of $T$ and disorder.
This result has also been verified by Monte Carlo simulations.
Our numerical results show that the temperature region where the $T=0$ behavior
dominates is narrower than the crossover region.

In this paper the dynamics of one-dimensional CDWs in the 
creep regime is also studied.
According to the scaling theory for manifolds \cite{Natter90}, in the case
of the CDWs the dependence of
 the creep velocity $v$, on external an electric field $E$,
is given by the following expression
\begin{equation}
v(E) \; \sim \; \exp\left[ -\frac{T_{\xi}}{T}\left( \frac{E_{\xi}}{E} 
\right)^{\mu} \right]
 \; , \quad \mu \; = \; \frac{d-2}{2} \; .
\label{eq.0}
\end{equation}
Here $T_{\xi}$ and $E_{\xi}$ are parameters which depend on the effective barrier height and the typical length scale of the problem\cite{NatSchei}. 
Formula (\ref{eq.0}) is valid for $d > 2$ for which the exponent $\mu$ is positive
(for $d=2$ one has a logarithmic behavior and the exponential is replaced
by a power law\cite{Shapir92}). However, it is not longer valid for $d<2$ as $\mu$ becomes negative. 
Thus, the problem of creep dynamics of CDWs in one-dimensional
systems remains open. 

In this paper we have made an attempt to develop the theory for creep dynamics
in one dimension. We have shown that
the dependence of the creep velocity on the electric field is given by the hyperbolic
sine function. Such behavior is in sharp contrast to the 
non-analytic behavior (\ref{eq.0}) for $d \ge 2$. 
As will be demonstrated below, the analytic dependence is due to the absence of
the transverse motion of manifolds. 
The analytical prediction was confirmed by solving the corresponding Langevin
equations numerically.
Our results seem to be in agreemnt with the 
experimental findings of Zaitsev-Zotov
\cite{Zaitsev93} for the temperature dependence of the current at 
low values of $T$ (see Fig. 2 in Ref. [\onlinecite{Zaitsev93}]).

\section{Model and phase-phase correlation function}\label{sec:model}
The charge--density $\rho(x)$ of a 1D CDW can be expressed as
 $  \rho(x)=\rho_0+\rho_1\cos{\big(Qx+\varphi(x)\big)}$
where $Q=2k_F$ denotes the wave vector of the undistorted wave, 
$k_F$ the Fermi--momentum and $\varphi(x)$ a slowly varying phase variable.
The Hamiltonian of the phase field is then given by~\cite{Gorkov77}
   \begin{eqnarray}
   {\cal H}&=&\int dx\,\left\{  \frac{1}{2}\hbar v_F
   \left(\frac{\partial}{\partial x}\varphi\right)^2 \right.\nonumber\\
   &&-\left. \sum\limits_i V_i\delta(x-x_i)\cos{(\varphi+Qx)}\right\}\,,
   \label{eq.2}
   \end{eqnarray}
where $V_i>0$ and $x_i$ denote the strength and the position of the 
impurity potential acting on the CDW. The mean impurity distance $1/c$
is assumed to be large in comparison with the CDW--wave length such
that $Q\gg c$. 

For the further treatment it is important to separate between the cases
of weak and strong disorder, respectively. 

For \textit{weak disorder} $V_i\ll\hbar v_F c$ the Fukuyama--Lee length~\cite{FuLee77}
$L_c=(\hbar v_F^{\phantom{1}})^{2/3}(c\overline{V_i^2})^{-1/3}$
is large compared with the impurity
distance. Here $\overline{V_i^2}$ denotes the averaged potential 
strength of the impurity. 
In the following we will therefore restrict ourselves to the case
   \begin{equation}
   L_c\gg c^{-1}\gg Q^{-1}\,.
   \label{eq.3}
   \end{equation}
The length scale $L_c$ sets an energy scale
   \begin{equation}
   T^{\ast}=\left(\hbar v_F^{\phantom{1}}c\overline{V^2_i}\right)^{1/3}=
   \hbar v_F^{\phantom{1}}L_c^{-1}\,.
   \label{eq.4}
   \end{equation}
Typically $T^{\ast}/T$ if of the order $10^3-10^7$ [\onlinecite{MiddFi91}].

Under condition (\ref{eq.3}) the Hamiltonian can be 
rewritten in the form of the random field XY--model: 
   \begin{eqnarray}
   {\cal H}&=\int dx\Bigg\{&\frac{1}{2}\hbar v_F
   \left(\frac{\partial}{\partial x}\varphi-g(x)\right)^2\nonumber\\
   &&-V\cos{\big(\varphi-\alpha(x)\big)}\Bigg\}\,,
   \label{eq.5}
   \end{eqnarray}
where $V^2_{\phantom{1}}=\overline{V^2_i}c$. $\alpha(x)$ is a random phase with
zero average and
   \begin{equation}
   \overline{e^{i\big(\alpha(x)-\alpha(x^{\prime})\big)}}=
   \delta(x-x^{\prime})\,.
   \label{eq.6}
   \end{equation}
In (\ref{eq.5}) we added also a linear gradient term which in
general will be generated under a renormalization group transformation.
Here
   \begin{equation}
   \overline{g(x)}=0\quad \; , \; \quad \overline{g(x)\,g(x^{\prime})}=
   \sigma\delta(x-x^{\prime})\,.
   \label{eq.7}
   \end{equation}
Model (\ref{eq.5}) exhibits a statistical tilt symmetry 
\cite{Schultz88},
which excludes a renormalization of the stiffness constant $\hbar v_F$.
This can most easily be seen from the replica Hamiltonian
corresponding to (\ref{eq.5})
   \begin{eqnarray}
   {\cal H}_n&=&\sum\limits_{\alpha,\beta}\int\limits_0^Ldx\,\left\{
   \frac{1}{2}\hbar v_F\left(\frac{\partial}{\partial x}\varphi_{\alpha}
   \right)^2\delta_{\alpha\beta}\right.\nonumber\\ 
   &&- \frac{(\hbar v_F)^2\sigma}{2T}\left(\frac{\partial}{\partial x}\varphi_{\alpha}\right)
   \left(\frac{\partial}{\partial x}\varphi_{\beta}\right)\\
   &&-\left.\frac{V^2}{4T}\cos{(\varphi_{\alpha}-\varphi_{\beta})}\right\}\,.\nonumber
   \label{eq.8}
   \end{eqnarray}
Adding a term $-g_0\int_0^L(\partial\varphi/\partial x)dx$ to
${\cal H}$, the full stiffness constant $\hbar\tilde v_F$ follows from the 
average free energy by
   \begin{equation}
   (\hbar\tilde v_F)^{-1}=-\frac{1}{L}
   \frac{\partial^2\bar F}{\partial g^2_0}\Big|_{g_0=0}\,.
   \label{eq.9}
   \end{equation}
Rewriting ${\cal H}_n$ in terms of $\tilde\varphi(x)=\varphi(x)-
\frac{g_0}{\hbar v_F}x$, the only remaning $g_0$--term is 
$-\frac{1}{2}(g_0^2/\hbar v_F)$, which proves our statement.

For the further discussion it is convenient to go over to
rescaled length and energy units. With $x=L_cy$ and
$\varphi(x)=\tilde\varphi(y)$
   \begin{equation}
   \frac{\cal H}{T}=\frac{1}{\tilde T}\int dy\left[\frac{1}{2}\left(
   \frac{\partial\tilde\varphi}{\partial y}-\tilde g(y)\right)^2+
   \cos{\big(\tilde\varphi-\tilde\alpha(y)\big)}\right]\,,
   \label{eq.10}
   \end{equation}
where $\tilde T=T/T^{\ast}$ with $T^{\ast}$ defined in (\ref{eq.4})
and
   \begin{eqnarray}
   \overline{e^{i\big(\tilde\alpha(y)-\tilde\alpha(y^{\prime})\big) } }&=&
   \delta(y-y^{\prime})
   \overline{\tilde g(y)\,\tilde g(y^{\prime})}\nonumber\\ 
    &=&\tilde\sigma\delta(y-y^{\prime})\,,\quad\tilde\sigma=L_c\sigma\,.
   \label{eq.11}
   \end{eqnarray}
Since $\sigma=0$ in the initial Hamiltonian (\ref{eq.2}), the
static properties of the model are characterized by $L_c$ and $T^{\ast}$.

Defining
\begin{equation}
\frac{\partial\tilde\phi}{\partial y} \; \; \equiv \; \;
\frac{\partial\tilde\varphi}{\partial y}-\tilde g(y)
\label{eq.12a}
\end{equation}
we have
\begin{equation}
\tilde{\varphi}(y) \; \; = \; \; \tilde{\phi}(y)-\tilde\phi(0)
+ \int_0^y \, \tilde{g}(z) dz \; .
\label{eq.12b}
\end{equation}
To the lowest order of  perturbation theory one can
ignore the non--linear term in (\ref{eq.10}). Then, using (\ref{eq.12b})
and (\ref{eq.7}), we
obtain for the pair correlation function
   \begin{eqnarray}
   C(x) &=& \big<\overline{\big(\varphi(x)-\varphi(0)\big)^2}\big>
   = \big<\overline{\big(\tilde{\varphi}(y)-
     \tilde{\varphi}(0)\big)^2}\big> \nonumber\\ 
   &=& \big<\big(\tilde\phi(y)-\tilde\phi(0)\big)^2\big>+
       \int_0^y \, \int_0^y \,dz\,dz^{\prime}\overline{\tilde g(z)\tilde g(z^{\prime})} \nonumber \\
   &=& \left(\tilde T+\tilde \sigma\right)|y|=
       \left(\frac{T}{\hbar v_F}+\sigma\right)|x|^{2\zeta}\,.
   \label{eq.12}
   \end{eqnarray}
with the roughness exponent $\zeta=\frac{1}{2}$. 

It should be noted that using a transfer matrix method
Feigel'man~\cite{Feigel80} 
found in the zero temperature limit a value of
$\sigma$ of the order $L_c^{-1}$, i.e. $\tilde\sigma$ is
of the order 1. The same result was obtained by Villain and 
Fernandez~\cite{ViFer84} by a real space renormalization group transformation.
Our formula (\ref{eq.12}) obtained by the renormalization of the replica
Hamiltonian is the extension of their result to the finite temperature
case.

To check the theoretical prediction
(\ref{eq.12}) numerically we have performed a detailed Monte 
Carlo study using the discretized version of Eq. (\ref{eq.10}) with $\tilde g(y)=0$. 
The free boundary condition is implemented.
The acceptance ratio of Metropolis moves was controlled to be around
0.5 for the whole run.
The equilibration is checked by monitoring the stability
of data against runs which are at least three-times longer. The first half of
Monte Carlo steps are not taken into account when averaging.

\begin{figure}
 \epsfig{file=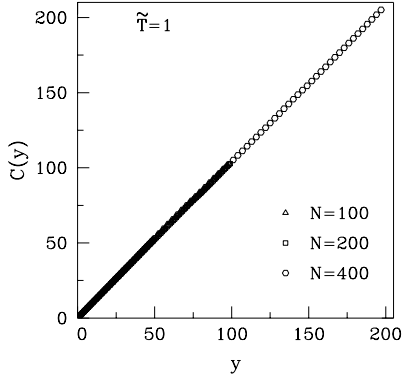,width=6cm}
\caption{\label{fig.1}
The distance dependence of the correlation function $C(y)$ for
various system sizes. $N=100, 200$ and 400, $\tilde T=1$. The results
are averaged over 50 -- 200 samples.}
\end{figure}

Fig.~\ref{fig.1} shows the pair correlation function $C(y)$ 
as a function of $y$ for different system sizes and for $\tilde T=1$.
Clearly, $C(y)$ has the same slope for any value of $N$.
In what follows we will take $N=200$ to calculate $C(y)/y$.

\begin{figure}
\epsfig{file=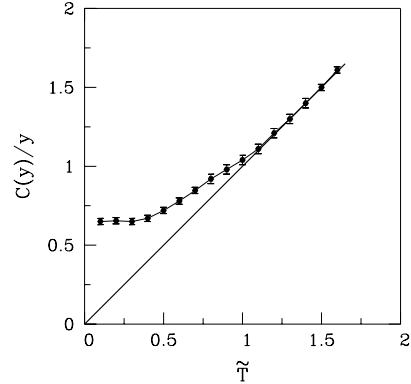,width=6cm}
\caption{\label{fig.2}
The temperature dependence the slope  $C(y)/y$.
The results are averaged over 50 -- 200 samples. The straight line corresponds to $C(y)/y=\tilde T$.}
\end{figure}

Fig.~\ref{fig.2} shows $C(y)/y$ as a function of $\tilde T$.
Above $\tilde{T}\approx 1.3$ the disorder is irrelevant and 
the high-temperature behaviour sets in.  For $\tilde{T} \le 0.3$
the disorder dominates and $C(y)/y$ becomes independent of $T$.
Clearly, our numerical results support prediction (\ref{eq.12}) for low and high temperatures. 
From Fig. 2 we obtain $\tilde{\sigma} \approx 0.65$ and it is of the
order of unity as predicted by theory.
 The crossover region $0.3 \le T \le 1.3$ is wider than the region
 where the $T=0$ behavior is valid ($T \le 0.3$).

\section{Dynamics}\label{sec:dynamics}

Since the roughness exponent is $\zeta=1/2$ the random field potential creates
 rugged energy barriers $\hbar v_F^{\phantom{1}}L^{-1}_c=
T^{\ast}_{\phantom{1}}$. Barriers on larger length scales are of 
the same order. In the following we will use these findings to
determine the creep motion of the CDW under the influence of an
external electric field $E$. To this aim we have to add
to the Hamiltonian (\ref{eq.5}) the following  term \cite{Gruener94}
\begin{equation}
H_{ext} \; \; = \; \; -\int dxE\varphi(x) \;
\label{eq.15a}
\end{equation} 

The equation of the (overdamped) CDW is then given by 
   \begin{equation}
   \frac{\partial\varphi}{\partial t}=-\gamma\left(
   \frac{\delta {\cal H}}{\delta\varphi}-E\right)+\eta(x,t)\,,
   \label{eq.15}
   \end{equation}
where $\gamma$ is a kinetic coefficient and $\eta(x,t)$ a Gaussian thermal
noise characterized by $\big<\eta\big>=0$ and
   \begin{equation}
   \big<\eta(x,t)\,\eta(x^{\prime},t^{\prime})\big>=
   2T\gamma\,\delta(x-x^{\prime})\,\delta(t-t^{\prime})\,.
   \label{eq.16}
   \end{equation}
The rescaling in the previous section amended by a rescaling of time 
according to $\tau=\gamma\frac{T^{\ast}}{L_c}t$
leads to an equation of motion which includes as the only parameters
$\tilde T=T/T^{\ast},\;\tilde\sigma=\sigma/L_c$
and $\tilde E=E/E^{\ast}$, where $E^{\ast}=T^{\ast}/L_c$ is of the order
of the $T=0$ depinning threshold field $E_T$.

For temperatures $T\gg T^{\ast}$ the energy landscape is essentially 
flat and the CDW makes a damped motion with $\dot\varphi\approx\gamma E$.
In the opposite case of $T\ll T^{\ast}$ energy barriers on the scale $L$ 
are of the order $E$ ($L>L_c$). In fact, in the absence of the external field 
the energy barrier $E_B$ is of the order of $T^{\ast}$. 
From (\ref{eq.15a}) it is clear that due to  the external field
the energy barrier has an additional term proportional to
\begin{equation}
EL \big<\varphi^2\big>^{1/2} \; \; \sim \; \; EL\left(\frac{L}{L_c}\right)^{1/2} \; .
\label{eq.17a}
\end{equation}
In obtaining the last equation the roughness exponent $\zeta=1/2$ 
was taken into account. So we have the following expression for the 
energy barrier
   \begin{equation}
   E_B^{\pm}(L)\approx c_BT^{\ast}\mp c_EE\left(\frac{L}{L_c}\right)^{1/2}L\,,
   \label{eq.17}
   \end{equation}
where $c_B$ and $c_E$ are constants of order unity and the $\pm$ sign refers to
the motion parallel and antiparallel to the external
field.

Since the largest energy barrier arise when $L\approx L_c$, we
find from the Arrhenius law for the creep velocity of the CDW
   \begin{eqnarray}
   v(E)\approx \frac{\gamma}{2}\frac{T}{L_c} \left( \exp(-E_B^+/T)-\exp(-E_B^-/T) \right) \nonumber\\
= \gamma\frac{T}{L_c}e^{-c_B(T^{\ast}/T)}
   \sinh{\frac{c_EEL_c}{T}}\,.
   \label{eq.18}
   \end{eqnarray}
In (\ref{eq.18}) we have chosen the prefactor in such a way that for
$T\gg T^{\ast}$ and $EL_c\ll T$ the linear behaviour $v\sim\gamma E$
is recovered.
So the creep law shows the conventional Kim--Anderson behaviour and
is drastically different from the behaviour in higher dimensions
where a non--analytic dependence of $v$ on $E$ is found~\cite{Natter90}.

We now try to understand the difference between $d=1$ and $d>1$ systems
qualitatively. For the creep motion of a vortex lattice or a CDW in
higher than one dimension, strings (or manifolds) must get thermally activated
over a potential barrier of a typical length $L_z$ \cite{NatSchei}. 
In the case of a
vortex lattice the strings are vortices whereas for a CDW it is a
phase manifold of dimensionality $d-1$.
Here two factors should be taken into account. First, due to the disorder
and the roughness of manifolds,
the height of the barriers scales as a power law in $L_z$
with the exponent
of the free energy fluctuations . Strickly speaking, this is
correct for the case without a driving force. However, in the creep regime
where the force is much smaller than  the depinning threshold
such an approximation is still valid. 
Second, in the presence of a weak driving force
the strings try to move a long distance to overcome the barrier.
When the manifold overcomes such a barrier, one gains an energy which
is proportional to $E$ and is generally a non-analytic function of $L_z$.
The interplay between the energy gain and the existence of barriers
leads to a non-trivial dependence of $L_z$ on $E$ or to the non-analytic
dependence of the creep velocity on the driving force.
Our discussion is based on the review of Nattermann and Scheidl~[\onlinecite{NatSchei}].
A similar argument may be also found in the review of Blatter {\em et al.}~[\onlinecite{Blatter}].

In the $d=1$ case the phase manifolds are points and one has no strings 
perpendicular to the direction of the creep motion. Due to the lack
of the transverse motion of strings, the typical height of barriers
is defined by the disorder strength, which is proportional to $T^*$.
The driving force changes the barriers by an amount given
by Eq. (\ref{eq.17}). So the crucial difference between $d=1$ and $d>1$ systems
is that the latter have a transverse motion of strings leading
to the non-analytic dependence of $v(E)$ on $E$.

In the next section we will check the prediction (\ref{eq.18}) for the creep motion by
a numerical simulation.

\section{Creep Simulation}

\begin{figure}
 \epsfig{file=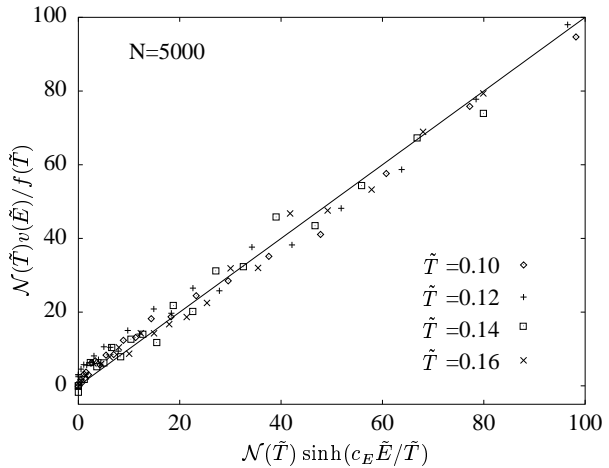,width=8cm}
\vspace{0.4cm}
\caption{
\label{fig.3}
The driving force dependence of the creep-velocity for fixed $\tilde T$. $f(\tilde T)=\gamma\frac{T}{L_c} e^{-\frac{c_B}{\tilde T}}$ 
and ${\cal N} (\tilde T)$ is a scaling-factor for a better illustration of the linear behaviour with values: 
${\cal N} (0.10)=1.0$, ${\cal N} (0.12)=2.5$, ${\cal N} (0.14)=6.5$ and ${\cal N}(0.16)=9.0$.}
\end{figure}

In our simulation we used the following discrete and rescaled version of~(\ref{eq.15})
 \begin{eqnarray}
  \frac{\Delta \tilde{\varphi}_i}{\Delta\tau}=&&(\tilde{\varphi}_{i+1}-2\tilde{\varphi}_i+\tilde{\varphi}_{i-1})\nonumber\\
  &&+\sin(\tilde{\varphi_i}-\tilde{\alpha_i})+\tilde E+\tilde\eta(i,\tau)\,,\quad i=1\ldots N
 \end{eqnarray}
The first term is the one-dimensional lattice Laplacian and $\tilde{\alpha_i}$ is uniformly distributed in the interval $[0,2\pi[$. 
The thermal noise $\tilde\eta(i,\tau)$ is defined by Eq. (\ref{eq.16}).

The creep velocity $v$ is given by
 \begin{equation}
  v(\tilde E,\tilde T)=\left\langle\frac{1}{N}\sum_{i=1}^{N}{\frac{\Delta \tilde{\varphi_i}}{\Delta\tau}}\right\rangle_{\tau} \; .
 \end{equation}
It should be noted that $v\propto j_{cdw}$, where $j_{cdw}$ is the CDW-current.
The equation of motion is integrated by a 
\textit{modified Runge-Kutta} algorithm suitable for 
stochastic systems~\cite{Greenside81}. 
Periodic boundary conditions are applied.

We first tested our algorithm by studying the depinning transition
at zero temperature. We found a threshold 
field of $\tilde E_T\approx 0.22$ with a critical exponent $\xi=0.57\pm0.07$.
This value of $\xi$ is in agreement with previous works 
~\cite{SibaniLitt90,MyersSethna93}.

For the creep simulations we used a time of 
$1000\tau_0$ and $\Delta\tau=0.05\tau_0$.
Runs which are three-times longer do not change the results in any substantial way.
In order to check the predicted behaviour (\ref{eq.18}) we first fix $\tilde T$
and vary $\tilde E$. In this case
we took a system size of $5000$ and the results were averaged over 50 
disorder realizations
(for larger system sizes the results remain almost the same). 
Our results are shown in Fig.~\ref{fig.3}. The linear fit by a straight line
in this figure indicates that Eq. (\ref{eq.18}) captures the field dependence
correctly.
By an iterative least-square fitting, we found $c_E=2.5\pm0.2$.

\begin{figure}
 \epsfig{file=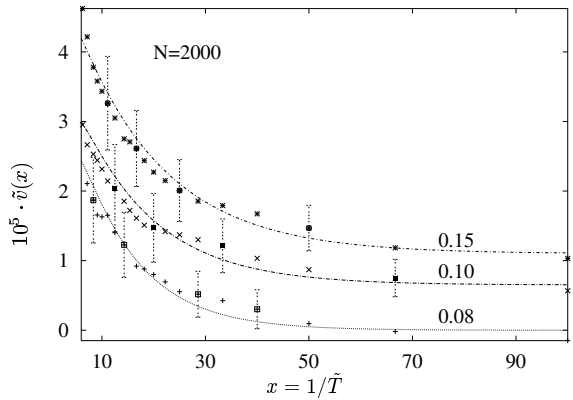,width=8cm}
\vspace{0.4cm}
\caption{
\label{fig.4}
The temperature dependence of the creep-velocity for different values of $\tilde E$
which are shown next to the curves.
For a better visualization we rescaled the data. For $\tilde E=0.08$, 
$3\tilde v(x)$ is plotted. For $\tilde E=0.10$ and $0.15$ the
data are shifted by $6\cdot 10^{-6}$ and $10^{-5}$, respectively. Typical
error bars are shown.}
\end{figure}

We now study the temperature dependence of $v(\tilde E)$ for fixed values of $\tilde E$.
We took $N=2000$ and averaged typically over $500$ samples.
The results are shown in Fig.~\ref{fig.4}.

Using $c_E=2.5$, we fitted the function 
$\tilde v(x)=\gamma e^{-c_Bx}\sinh(Bx)$ with $x=1/\tilde T$ and 
$B=c_E\tilde E$ and found $c_B=0.35\pm0.10$.
Combining the results shown in Fig. 3 and Fig. 4 one can see that
our simulation supports the predicted behaviour~(\ref{eq.18}) for
the creep-velocity.

It is clear from Fig. 4 that for fixed values of $\tilde E$ the current saturates at
low temperatures. This agrees with the experimental data presented in Fig. 2 of
Ref. \onlinecite{Zaitsev93}.  
At high temperatures the slopes of the current,
when plotted against $1/{\tilde T}$, were found to depend on $\tilde E$ \cite{Zaitsev93}. 
On the other hand,
as one can see from Eq. (3.4), the slopes should be independent of
$\tilde E$. Such discrepancy is due to the fact that formula (3.4) is valid
only for the creep regime.

In conclusion we have shown that although the glass behavior is governed by
the $T=0$ fixed point, the disorder has a dramatic effect on the low
temperature properties of 1D CDW's. At low temperatures
$C(y)/y$ is determimed by the disorder strength. In one dimension
the dependence
of the creep motion velocity on the driving field was found to be not
a non-analytic function
as in higher dimensions but an analytic one.
Our theoretical predictions were confirmed by numerical simulations.

\section*{Acknowledgments}

We are very grateful to T. Natterman for much useful advice and encouragement.
Discussions with J. Kierfeld, D. Stauffer and S. V. Zaitsev-Zotov are also acknowledged. 
One of us
(MSL) was supported in part by the Polish agency KBN (Grant number 2P03B-146-18).

\end{document}